\documentclass[aps,prl,superscriptaddress,twocolumn,showpacs]{revtex4-1}
\usepackage{graphicx}
\usepackage{dcolumn}
\usepackage{amsmath}
\usepackage{color}

\begin{document}

\title{Light-matter interaction in antiferromagnets:\\ the exchange-induced magnetic dipole mechanism}

\author{M.~V.~Eremin}
\author{M.~A.~Fayzullin}
\affiliation{Institute for Physics, Kazan (Volga region) Federal University, 430008 Kazan, Russia}
\author{Ch.~Kant}
\affiliation{Institut f\"ur Festk\"orperphysik, Technische
Universit\"at Wien, 1040~Wien, Austria}
\author{ A.~Loidl}
\author{J.~Deisenhofer}
\affiliation{Experimentalphysik V, Center for Electronic Correlations and Magnetism,
Institute for Physics, Augsburg University, D-86135 Augsburg, Germany}
\date{\today}

\begin{abstract}
We propose a novel mechanism for exchange-induced exciton-magnon
absorptions via hopping between two antiferromagnetically coupled
sites and a simultaneous magnetic dipole transition to an excited
orbital state. The obtained selection rules correspond to ones for
magnetic dipole transitions and are in agreement with the
exciton-magnon transitions observed in the quasi one-dimensional
Heisenberg antiferromagnet KCuF$_3$. The calculated magnon density
of states in combination with a structure dependent factor
identifies the observed optical magnon sideband to originate from
transverse magnon modes.

\end{abstract}

\pacs{78.20.Bh, 71.70.Ej, 75.30.Et, 75.30.Ds, 78.20.Ls}

\maketitle

The interaction of light and matter is one of the most fascinating
research fields in physics. In materials with coupled electronic,
lattice, and spin degrees of freedom such as, for example,
antiferromagnetic transition metal compounds, electric-dipole active
magneto-optical excitations including two-magnon, exciton-magnon,
and exciton-exciton absorptions have been observed
\cite{Halley1965,Greene1965,Ferguson1968}. In pioneering works
Tanabe and coworkers proposed mechanisms to explain these absorption
processes in terms of cooperative excitations of
antiferromagnetically coupled ions \cite{Tanabe1965,Tanabe1966,
Tanabe1967,Shinagawa1971,Fujiwara1972}. The observation of
electric-dipole active three-magnon in $\alpha$-Fe$_2$O$_3$
\cite{Azuma2005,Tanabe2005a,Tanabe2005b} and one-magnon processes
(electromagnons) in multiferroic manganites, respectively, pushed
the search for such excitations and their microscopic mechanisms
\cite{Pimenov2006,Mostovoy2006}.

\begin{figure}[t]
\includegraphics[width=59mm,clip]{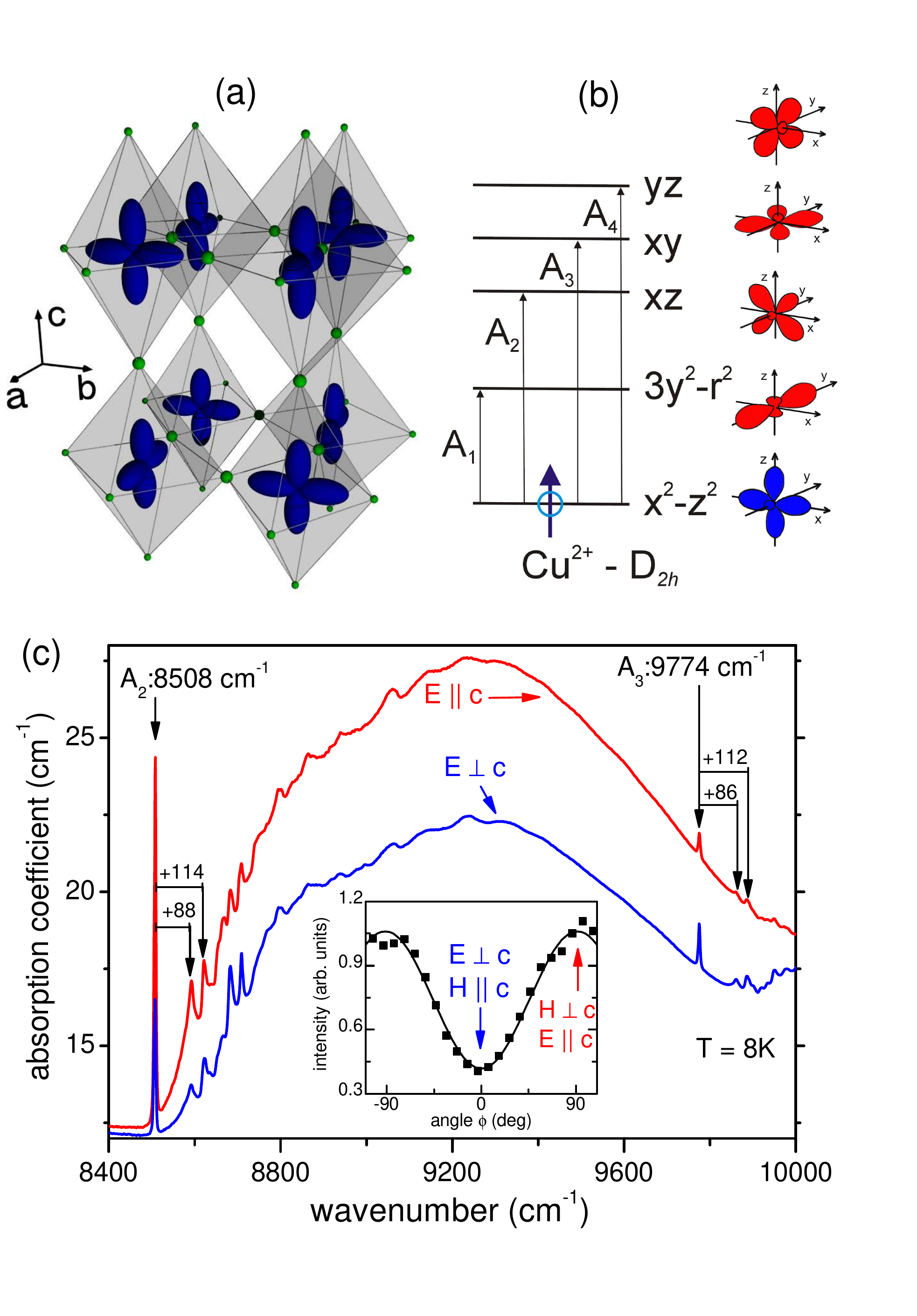}
\caption{\label{fig:structure}(a) Schematic view of the alternating
orbital ordering in KCuF$_3$. (b) Splitting of the $d$-levels of the
Cu$^{2+}$ ions in $D_{2h}$ symmetry and observed excitations
$A_1-A_4$. (c) Optical absorption spectra focussing at the onset of
$A_2$ and $A_3$. Peaks with an energy separation of 88~cm$^{-1}$ and
114~cm$^{-1}$ to the purely excitonic lines are indicated. Inset:
polarization dependence of the exciton line of $A_2$ (see
Ref.\onlinecite{Deisenhofer2008})}
\end{figure}

Recently, the observation of magnon sidebands at the onset of the
phonon assisted orbital transitions was reported in the quasi-one
dimensional Heisenberg antiferromagnet KCuF$_3$
\cite{Deisenhofer2008}. The structure of this compound is sketched
in  Fig.~\ref{fig:structure}(a) together with the antiferroorbital
arrangement of the $d_{x^2-z^2}$ and $d_{y^2-z^2}$ orbital states of
the Cu$^{2+}$ ions with a $3d^9$ electronic configuration. In
Fig.~\ref{fig:structure}(b) we show the splitting of the $d$-levels
at the copper site in $D_{2h}$ symmetry (considering one hole
occupying a $d_{x^2-z^2}$ state) and indicate the excitations
$A_1$-$A_4$ observed in \cite{Deisenhofer2008}.

It is important to note that transitions to the excited $d_{xz}$,
$d_{xy}$, and $d_{yz}$ orbital states are magnetic dipole allowed
via the orbital momentum operators $l_\alpha$ ($\alpha=y,z,x$),
respectively. In Fig.~\ref{fig:structure}(c) the absorption spectra
at 8~K for polarization of the incoming light $\textbf{E}\parallel
c$ and $\textbf{E}\perp c$ are shown, focussing on the onset of
$A_2$ and $A_3$. The sharp peaks at 8508~cm$^{-1}$ and
9775~cm$^{-1}$ correspond to the respective purely orbital
transitions with energy, while the sidebands at a distance of about
88~cm$^{-1}$ and 114~cm$^{-1}$ were identified as exciton-magnon
transitions \cite{Deisenhofer2008}. The inset of
Fig.~\ref{fig:structure}(c) shows the polarization dependence of the
intensity of the peak at 8508 cm$^{-1}$ following a $\sin^2\phi$
dependence, with $\phi$ being the angle between the polarization and
the crystal axes. This polarization dependence of the zero-magnon
line and the whole set of sidebands is in agreement with the
selection rules expected for magnetic dipole transitions, but at
odds with the expected behavior from the canonical exchange-induced
electric dipole (EIED) mechanism.

Here we propose a novel mechanism to explain the exciton-magnon
transitions in KCuF$_3$. The resulting exchange-induced magnetic
dipole (EIMD) processes allow to describe the observed polarization
dependence and are compared to the EIED mechanism.

We consider two exchange-coupled magnetic ions at sites \textit{a}
and \textit{b}. Using the canonical transformation with S-matrix
formalism, the effective Hamiltonian
\begin{equation}
 H_{eff}=F+[F,S]+\frac{1}{2}[[F,S],S]+...\label{eq1}
\end{equation}
is obtained and accounts for all possible processes up to third
order of perturbation theory. The operator \textit{F} describes
transitions between different orbital states $\varphi$ and
$\varphi^\prime$ of the ion at site \textit{a}
\begin{equation}
 F=\mu_B\sum_{\varphi,\varphi^\prime}a_{\varphi}^+<\varphi|H_a^\alpha l_a^\alpha|\varphi^\prime>a_{\varphi^\prime}. \label{eq2}
\end{equation}
Here, $\mu_B$ denotes the Bohr magneton, $H_a^\alpha$ the magnetic
component of the electromagnetic wave along $\alpha=x,y,z$ and
$l_a^\alpha$ is the orbital momentum operator at site \textit{a}.
$a_{\varphi}^+$ and $a_{\varphi^\prime}$ are the electron creation
and annihilation operators. The operator \textit{S} is given by
\begin{equation}
 S=\sum_{\xi,\eta\prime}\frac{t_{\xi\eta\prime}}{|\Delta_{\xi\eta\prime}|}b^+_\xi a_{\eta\prime}-\sum_{\eta,\xi\prime}\frac{t_{\eta\xi\prime}}{|\Delta_{\eta\xi\prime}|}a^+_\eta b_{\xi\prime},
\label{eq3}
\end{equation}
where $t_{\xi\eta\prime}$, $t_{\eta\xi\prime}$  are effective
hopping integrals and $\Delta_{\xi\eta\prime}$,
$\Delta_{\eta\xi\prime}$ are charge-transfer energies. The
commutator $[F,S]$ describes processes of the electron hopping
between the two sites with simultaneous excitation by light
absorption as well as emission from excited states of the
\textit{a}-site ion. Note that none of these processes involves a
spin flip.

\begin{figure}[t]
\includegraphics[width=95mm,clip, angle=90]{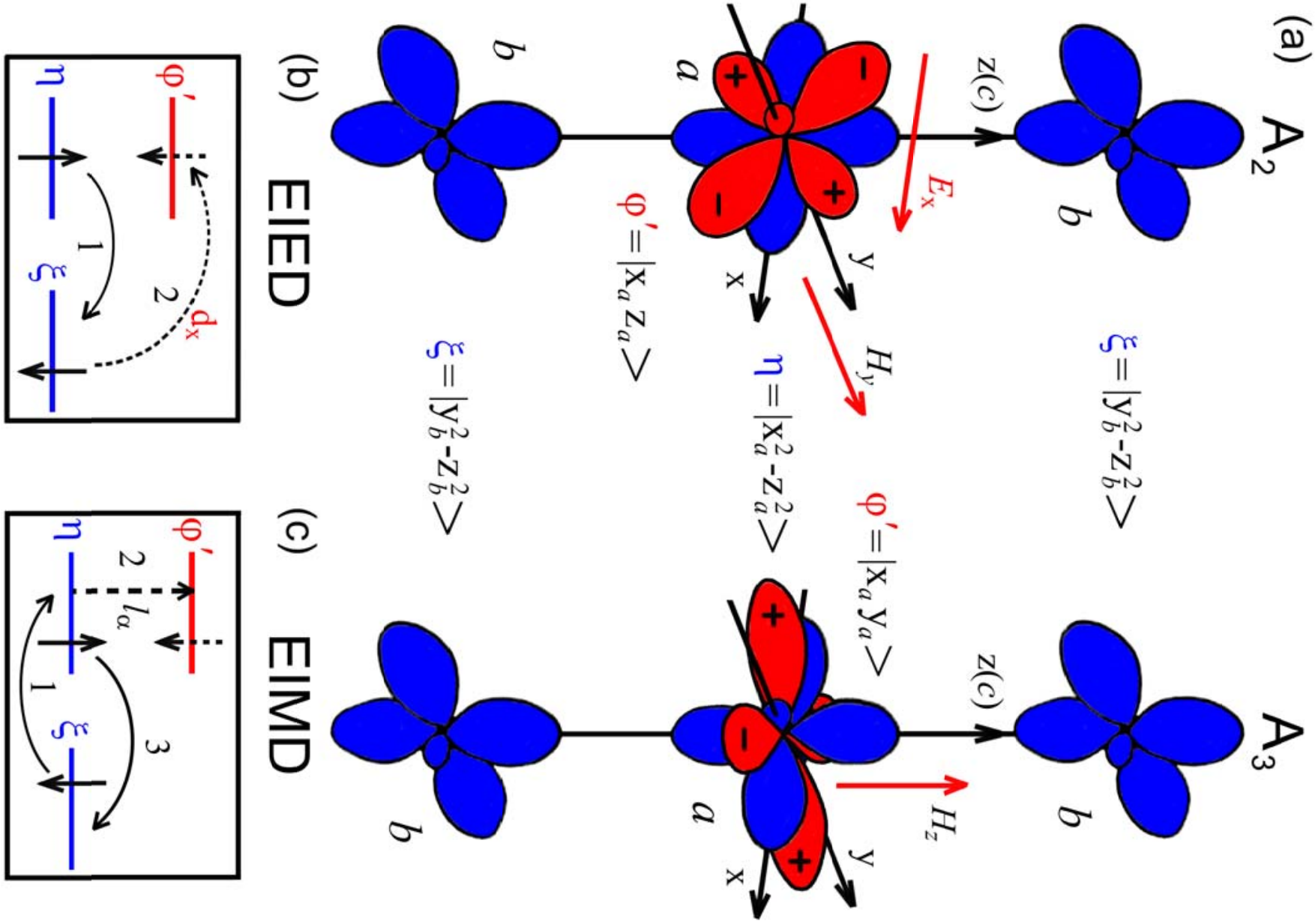}
\caption{\label{fig:distribution}(a) Sketch of the spatial charge
distributions of ground and excited orbitals of copper ions in
KCuF$_3$ along the antiferromagnetic \textit{c}-axis for the
transitions $A_2$ (left) and $A_3$ (right) and the corresponding
components of the electric and magnetic radiation field yielding
non-zero matrix elements for the EIMD and EIED mechanisms. (b)
Schematic plot of the involved processes for the EIED mechanism and
(c) for the novel EIMD mechanism.}
\end{figure}

One of the virtual processes we are considering here is sketched in
Fig.~\ref{fig:distribution}(c). The process involves a magnetic
dipole transition to an excited orbital state at site \textit{a}
(dashed arrow) and hopping processes of holes between the
 \textit{a} and \textit{b} sites in the ground states (solid
arrows). The effective Hamiltonian corresponding to the relevant
processes of the magnetic dipole transition can be derived from the
double commutator in Eq.~(\ref{eq1}). Here we are interested in the
effective operator for the absorption process which can be written
in terms of spin operators for the exchange-coupled ions at the
\textit{a} and \textit{b}-sites

\begin{equation}
H_{eff}^{\varphi\to\varphi\prime}=\frac{\mu_B}{2\Delta}<\varphi|H_a^\alpha
l_a^\alpha|\varphi\prime>J^c_{ab}[\mathbf{\sigma}_a\textbf{S}_b-1/4],\label{eq4}
\end{equation}
where $J^c_{ab}=2t_{\eta\xi}^2/|\Delta_{\eta\xi}|$  is the
superexchange coupling constant between the ground states of ions
\textit{a} and \textit{b}. This contribution has to be summed over
all nearest neighbors at sites \textit{b}. Strictly speaking the
introduced operator $\mathbf{\sigma}_{a}$ is not a true spin
operator, because it has matrix elements between the different
orbital states. Evidently, the suggested EIMD process is only
effective for a pair of antiferromagnetically coupled ions. This
effective Hamiltonian can be interpreted as follows: the absorption
(or creation of an exciton) on considering an optically active
center on site \textit{a} gives rise to magnons in the surrounding
magnetic sublattice.

Let us now discuss this mechanism for the case of KCuF$_3$ and
compare it to the EIED mechanism suggested by Tanabe, Moriya and
Sugano \cite{Tanabe1965}. In Fig.~\ref{fig:distribution}(a) we
schematically show spatial distributions of ground and excited
orbital states of copper ions along the antiferromagnetic
\textit{c}-axis. The scheme on the left hand side refers to the
A$_2$ transition and the one on the right to the A$_3$ transition.
The respective ground state orbitals at sites $a$ and $b$ are
$\eta=|x_a^2-z_a^2\rangle$ and $\xi=|y_b^2-z_b^2\rangle$, the
respective excited orbital states at site $a$ corresponding to $A_2$
and $A_3$ are $\varphi^\prime=|xz\rangle$ and $|xy\rangle$.

The case of the EIED mechanism suggested by Tanabe, Moriya and
Sugano \cite{Tanabe1965} is illustrated in
Fig.~\ref{fig:distribution}(b).  The up-spin at the \textit{a} site
($\eta$) hops to the \textit{b} site ($\xi$) while the down-spin at
site \textit{b} is transferred to an excited orbital state at site
\textit{a} due to interaction of the electric-dipole moment with the
electric-field component of the incoming light. This mechanism will
only lead to a non-zero matrix element for the A$_2$ transition if
$\textbf{E}\perp c$. In case of the A$_3$ transition the EIED
transition there is no non-zero matrix element. These selection
rules, however, are at odds with the experimentally observed
polarization dependence.

The processes involved in the EIMD mechanism are schematically
sketched in Fig.~\ref{fig:distribution}(c). The down-spin at site
\textit{b} (orbital $\xi$) hops to the ground state of ion
\textit{a}, and then is excited to the orbital state $\varphi\prime$
due to the interaction of the orbital momentum with the magnetic
field component of the incoming light and simultaneously the up-spin
at site \textit{a} hops to site \textit{b}. The selection rules,
therefore, correspond to the ones for magnetic dipole transitions in
agreement with the experimental observation \cite{Deisenhofer2008}.

\begin{figure}[t]
\includegraphics[width=90mm,clip]{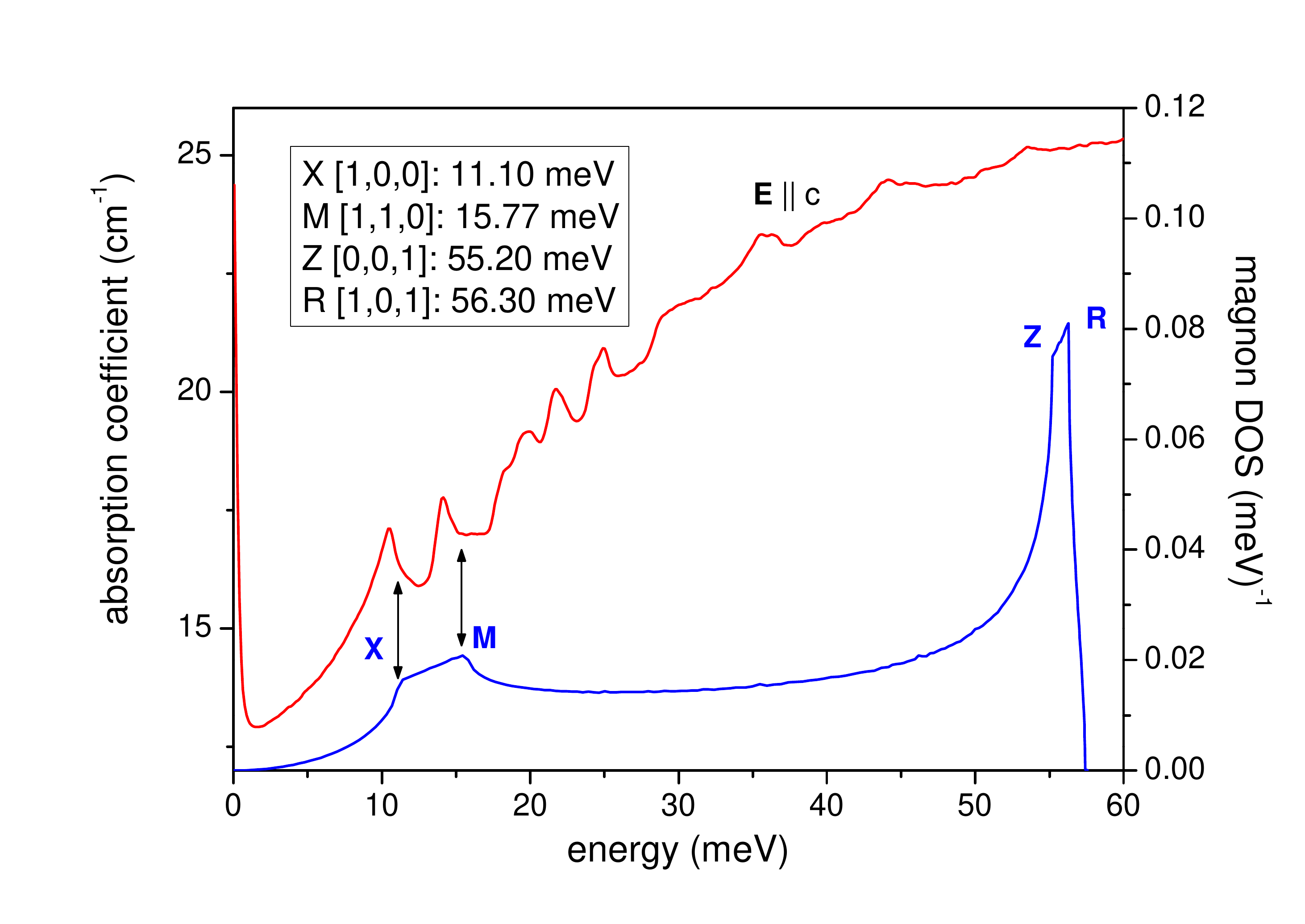}
\caption{\label{figure:DOS}Absorption spectrum of the exciton-magnon
sidebands at 8 K for polarization $E\parallel c$ and calculated
magnon density of states in KCuF$_3$. The excitation energy of the
absorption spectrum corresponds to the distance to the zero-magnon
line at 8508 cm$^{-1}$. }
\end{figure}

Furthermore, we calculate the magnon density of states for KCuF$_3$
using the spin-wave Hamiltonian in the framework of a two-sublattice
model \cite{Satija1980}, but neglecting $XY$-like anisotropy and
contributions from the Dzyaloshinsky-Moriya interaction
\cite{Eremin2008}. Diagonalization via Holstein-Primakoff and
Bogoliubov transformations yields the magnon dispersion
\begin{equation}
E_k=\sqrt{{[8J^a(1-\theta_a)+2J^c]}^2-4{J^c}^2\gamma^2_c},
\label{eq5}
\end{equation}
where $J^c$, $J^a$ are the antiferro- and ferromagnetic exchange
coupling constants, respectively and $\theta_{a}=(\cos{k_{x}a}
+\cos{k_{y}a})/2$, $\gamma_{c}=\cos{k_{z}c}$. This dispersion was
used to fit the spin-wave spectrum observed in inelastic neutron
scattering experiments \cite{Satija1980} with $J^c=320$~K and
$J^a=1.6$~K. The same expression for the transverse magnon mode was
derived in \cite{Lake2005}. The magnon density of states (DOS)
$\rho(E)=\frac{1}{N}\sum_k^{BZ}\delta(E-E_k)$ was calculated by
using this dispersion and summing of the magnetic Brillouin zone
(BZ), which is doubled along the \textit{c} direction. The
calculated result is shown in Fig.~\ref{figure:DOS} together with
the optical absorption coefficient for light polarization
$E\parallel c$. The magnon DOS exhibits four singularities, which
corresponds to the points X$[\pi/a,0,0]$, M$[\pi/a,\pi/a,0]$,
Z$[0,0,\pi/2c]$ and R$[\pi/a,0,\pi/2c]$ in the magnetic BZ. The
corresponding energies are 11.1, 15.7, 55.2, and 56.3 meV,
respectively. The first two-magnon energies correspond very nicely
to the optically observed sideband peaks (see
Fig.~\ref{figure:DOS}). Although the DOS at the Z and R points is
very large with respect to X and M points, there are no prominent
features at these energies in the absorption spectra. This can be
explained by considering structure dependent factors which we
discuss in the following.

In general, the absorption coefficient caused by the magnetic field
component of light with frequency $\omega$ and polarization $\nu$ is
proportional to the imaginary part of the susceptibility and can be
evaluated in terms of Green's function \cite{Alexandrov}
\begin{equation}
A^\nu(E)\propto-E\frac{1}{V}\lim_{\delta\to0}\mathrm{Im}[G(M^*_\nu,M_\nu)_{E+i\delta}],
\label{eq7}
\end{equation}
where $V$ is the sample volume, $E=\hbar\omega$, $M_\nu$ is the
component of the transition magnetic dipole moment. Based on Eq.
\ref{eq4} in the low temperature approximation we define
\begin{equation}
M_\nu=\sum_{m,l}\pi^\nu_{ml}c^+_m S^+_l,
\label{eq8}
\end{equation}
where $c^+_m$ is a exciton creation operator of the spin-up
sublattice, $S^+_l$ is the spin operator of the spin-down sublattice
and
$\pi^\nu_{ml}=\mu_B<\varphi|l^\nu_a|\varphi\prime>J^c_{ml}/4\Delta$.

Following the Green's function formalism \cite{Parkinson1968} we
obtain
\begin{equation}
G(M^*_\nu,M_\nu)=\sum_k\frac{{|\pi^\nu_k |}^2 u^2_k}{E-E_{ex}-E_k}.
\label{eq9}
\end{equation}
Here, $E_k$ is the magnon energy, $E_{ex}$ is the exciton energy of
the zero-magnon (phonon) line,
\\$\pi^\nu_k=\mu_B
J^c\langle\varphi|l^\nu|\varphi\prime\rangle\cos{k_z}/2\Delta$, and
$u_k^2=(1+\sqrt{1+Z_k^2})/2$, $Z_k=2J^c\gamma_c/E_k$. Then
substituting the Green's function (Eq.~\ref{eq9}) to (Eq.~\ref{eq7})
we finally get
\begin{widetext}
\begin{equation}
A^\nu(E)\propto E\frac{\pi}{V}{\arrowvert J^c\frac{\mu_B}{2\Delta}\langle\varphi|l^\nu|\varphi\prime\rangle\arrowvert}^2\sum^{BZ}_k u^2_k\cos^2{k_z c}\cdot\delta(E-E_{ex}-E_k).
\label{eq10}
\end{equation}
\end{widetext}
The obtained absorption coefficient contains the structure-dependent
factor $\cos^2{k_z c}$, which filters out the singularities
corresponding to Z and R points in DOS during the summation over the
BZ, but the singularities corresponding to X and M points remain.
This explains the absence of prominent magnon sidebands at 55.2 and
56.3 meV. We would like to point out that in the case of the EIED
mechanism the factor $\cos^2{k_z c}$ is replaced by $\sin^2{k_z c}$
and, therefore, low-energy singularities corresponding to X and M
points in the magnon DOS are filtered out. The high-energy
singularities would be expected to a appear in the optical spectrum
for $\mathbf{E} \perp c$ as a result of the EIED mechanism and the
large DOS at the Z and R points, but no experimental evidence of
such sidebands are visible in the spectra. The reason of this
suppression of the allowed EIED is not clear at present. Thus, the
new EIMD mechanism allows to describe the observed low-energy magnon
sidebands at 11.1 and 15.7~meV in KCuF$_3$, but the further sideband
peaks visible in Fig.~\ref{fig:structure}(c) have no correspondence
in the calculated one-magnon DOS and their origin can not easily be
settled. Possibly, multi-magnon processes or the presence of spinons
will have to be considered and need further theoretical
investigations.

In summary, we have presented a new mechanism of a cooperative
exciton-magnon absorption in antiferromagnets -- the
exchange-induced magnetic-dipole mechanism. The selection rule
corresponds to the one for magnetic dipole transitions and describes
the experimental results in antiferromagnetic KCuF$_3$. Moreover, we
showed that for this novel mechanism the magnon density of states
and the structure factor in KCuF$_3$ are responsible for the
observability of low-energy magnon sidebands, while the high-enery
magnon sidebands are suppressed.

\begin{acknowledgments}
It is a pleasure to thank Sergei Nikitin for fruitful discussions.
We acknowledge support by DFG via TRR 80 (Augsburg-Munich). MVE is
partially supported by Ministry Education of the Russian Federation
via Grant No. 1.83.11.
\end{acknowledgments}

\end{document}